\def\be{\begin{equation}}
\def\ee{\end{equation}}
\def\ba{\begin{array}}
\def\ea{\end{array}}

\def\Cb{{\Bbb C}}

\documentclass[aps,amsmath,amssymb,amsfonts,showpacs,twocolumn,pra]{revtex4-1}
\usepackage{graphicx}
\usepackage{caption2}
\def\qed{\leavevmode\unskip\penalty9999 \hbox{}\nobreak\hfill
     \quad\hbox{\leavevmode  \hbox to.77778em{%
               \hfil\vrule   \vbox to.675em%
               {\hrule width.6em\vfil\hrule}\vrule\hfil}}
     \par\vskip3pt}

\newtheorem{theorem}{Theorem}
\newtheorem{corollary}{Corollary}

\begin{document}
\title{Lower Bound of Multipartite Concurrence Based on Sub-quantum State Decomposition}
\author{Xue-Na Zhu$^{1}$}
\author{Ming-Jing Zhao$^{2}$}
\author{Shao-Ming Fei$^{2,3}$}

\affiliation{$^1$Department of Mathematics, School of Science, South
China University of Technology, Guangzhou 510640, China\\
$^2$Max-Planck-Institute for Mathematics in the Sciences, 04103
Leipzig, Germany\\
$^3$School of Mathematical Sciences, Capital Normal University,
Beijing 100048, China}

\begin{abstract}

We study the entanglement of tripartite quantum states and provide
analytical lower bound of concurrence in terms of the concurrence of sub-states. The lower bound may improve all the existing lower bounds of concurrence. The approach is generalized to arbitrary dimensional multipartite systems.

\end{abstract}
\pacs{ 03.67.Mn, 03.65.Ud}
\maketitle

\section{Introduction}

Quantum entanglement \cite{s1} is considered to be the most
nonclassical manifestation of quantum mechanics  and plays an
important role not only in quantum information sciences
\cite{s2,s3,s4,s5} but also in condensed-matter physics \cite{Amico}.
The operational measure of entanglement for
arbitrary mixed states is not known yet, but the concurrence
\cite{s6,s7,s8,anote} is one of the well accepted entanglement measures
\cite{t1,t2,t3,t4,t5,t6}. Nevertheless, calculation of the concurrence is a formidable task
for higher dimensional case.

To estimate the concurrence
for general mixed states, efforts have been made toward the
analytical lower bounds of concurrence.
Therefore some nice algorithms and
progresses have been concentrated on possible lower bounds of the
concurrence for three quantum systems \cite{Xiu-hong
Gao.2006,Xiu-hong Gao.2008,Chang}. For arbitrary bipartite quantum
states, Ref. \cite{Yong} and Ref. \cite{zhao} provide a detailed proof of an
analytical lower bound of concurrence by decomposing the joint
Hilbert space into many $2 \otimes 2$ and $s \otimes t$-dimensional
subspaces, which may be used to improve all the known lower bounds
of concurrence. A natural problem is whether the arbitrary
dimensional tripartite quantum states can be dealed with this.

In this paper we provide a detailed proof of an analytical lower
bound of concurrence for tripartite quantum states by decomposing
the joint Hilbert space into any lower dimensional subspaces.
Moreover, the generalized lower bound of concurrence can be
generalized to the multipartite case.

\section{Lower bound of concurrence for tripartite quantum systems}

Let $H_{A_1} $, $H_{A_2}$, $H_{A_3} $ be three $N$-dimensional
Hilbert spaces associated with the systems $A_1$, $A_2$ and $A_3$. A
pure state $\vert\psi\rangle\in H_{A_1}\otimes H_{A_2}\otimes
H_{A_3}$ has the form
\begin{equation}
\vert\psi\rangle=\sum_{i=1}^{N}\sum_{j=1}^{N}\sum_{k=1}^{N}a_{ijk}\vert
ijk\rangle,
\end{equation}
where $a_{ijk}\in\Cb$,
$\sum_{ijk}|a_{ijk}|^2=1, \{\vert ijk\rangle\}$ is
the basis of $H_{A_1}\otimes H_{A_2}\otimes H_{A_3}$.

The concurrence of state ${\left\vert \psi \right\rangle }$ is
defined by, up to an $N$ dependent factor $\sqrt{{N}/{6(N-1)}}$,
\begin{equation}\label{cpure}
C({\left\vert \psi \right\rangle })=\sqrt{6-2\,\mbox{Tr}(\rho
_{A_1}^{2} +\rho_{A_2}^{2}+\rho _{A_3}^{2})},
\end{equation}
where the reduced density matrix $\rho_{A_1}$ (resp. $\rho _{A_2}$,
$\rho _{A_3}$) is obtained by tracing over the subsystems $A_2$ and
$A_3$ (resp. $A_1$ and $A_3$, $A_1$ and $A_2$). $C({\left\vert \psi
\right\rangle })$ can be equivalently written as \cite{anote}
\begin{widetext}
\be
\label{pure multi concurrence1}
C(\vert\psi\rangle)=\displaystyle\sqrt{\sum (\vert a_{ijk}a_{pqm}-a_{ijm}a_{pqk}\vert^2+
\vert a_{ijk}a_{pqm}-a_{iqk}a_{pjm}\vert^2+
\vert a_{ijk}a_{pqm}-a_{pjk}a_{iqm}\vert^2}).
\ee
\end{widetext}
The concurrence for a tripartite mixed state $\rho$ is defined by the convex roof,
\begin{equation}\label{e1}
C(\rho )\equiv \min_{\{p_{i},|\psi _{i}\rangle
\}}\sum_{i}p_{i}C({\left\vert \psi _{i}\right\rangle }),
\end{equation}
for all possible pure state decompositions $\rho
=\sum_{i}p_{i}|\psi_{i}\rangle\langle \psi _{i}|$, where
${\left\vert \psi_i \right\rangle } \in { H}_{A_1} \otimes
{ H}_{A_2}\otimes {H}_{A_3}$, $0\leq p_{i}\leq1$ and
$\sum_{i}p_{i}=1$.

To evaluate $C(\rho)$, we project high dimensional states to ``lower
dimensional" sub-states. For a given $N\otimes N \otimes N$ pure
state, we define its ``$m\otimes m\otimes m$", $m\leq N$, pure state
$|\psi\rangle_{m\otimes m\otimes
m}=\sum_{{i=i_1}}^{i_{m}}\sum_{{j=j_1}}^{j_{m}}\sum_{{k=k_1}}^{k_{m}}a_{ijk}
|ijk\rangle=B_1\otimes B_2 \otimes B_3|\psi\rangle$, where
$B_1=\sum_{{i=i_1}}^{i_m} |i\rangle\langle i|$,
$B_2=\sum_{{j=j_1}}^{j_m} |j\rangle\langle j|$,
$B_3=\sum_{{k=k_1}}^{k_m} |k\rangle\langle k|$. Its concurrence $C(\vert\psi\rangle_{m\otimes m\otimes
m})$ is similarly given by Eq.(\ref{pure multi concurrence1}), with the
subindices of $a$, $i$ (resp. $j$, $k$) associated with the system $A_1$ (resp.
$A_2$, $A_3$) running from $i_1$ (resp. $j_1$, $k_1$) to $i_m$
(resp. $j_m$, $k_m$).
In fact, for any $N\otimes N \otimes N$ pure state $|\psi\rangle$, there are ${N\choose m}^3$ different $m\otimes m\otimes m $ sub-states with respect to $|\psi\rangle$. Without causing confusion, in the following
we simply use $|\psi\rangle_{m\otimes m\otimes m}$ to denote one of such states, as these substates
will always be considered together.

Correspondingly for a mixed state $\rho$, we
define its ``$m\otimes m \otimes m$" mixed (unnormalized) sub-states
$\rho_{m\otimes m\otimes m}=B_1\otimes B_2 \otimes B_3 \rho
B_1^\dagger\otimes B_2^\dagger\otimes B_3^\dagger$. The concurrence of
$\rho_{m\otimes m \otimes m}$ is defined by $C(\rho_{m\otimes m
\otimes m})\equiv\min \sum_i p_i C(|\phi_i\rangle)$, minimized over all possible $m\otimes m \otimes m$
pure state decompositions of $\rho_{m\otimes m \otimes m}=\sum_i p_i
|\phi_i\rangle \langle \phi_i|$, with $\sum_i
p_i=tr(\rho_{m\otimes m \otimes m})$. The $m\otimes m \otimes m$
submatrices $\rho_{m\otimes m \otimes m}$ have the following form,
\begin{widetext}
\begin{eqnarray}
 \rho_{m\otimes m \otimes m}=\begin{pmatrix}
   \rho_{i_1j_1k_1,i_1j_1k_1}& ... & \rho_{i_1j_1k_1,i_1j_1k_m} & \rho_{i_1j_1k_1,i_1j_2k_1}&... & \rho_{i_1j_1k_1,i_mj_mk_m}
   \\
      \vdots & \vdots & \vdots & \vdots & \vdots&\vdots  \\
   \rho_{i_1j_1k_m,i_1j_1k_1}& ... & \rho_{i_1j_1k_m,i_1j_1k_m} & \rho_{i_1j_1k_m,i_1j_2k_1}&... & \rho_{i_1j_1k_m,i_mj_mk_m}
   \\
   \rho_{i_1j_2k_1,i_1j_1k_1}& ... & \rho_{i_1j_2k_1,i_1j_1k_m} & \rho_{i_1j_2k_1,i_1j_2k_1}&... & \rho_{i_1j_2k_1,i_mj_mk_m}  \\
      \vdots & \vdots & \vdots & \vdots & \vdots&\vdots  \\
   \rho_{i_mj_mk_1,i_1j_1k_1}& ... & \rho_{i_mj_mk_1,i_1j_1k_m} & \rho_{i_mj_mk_1,i_1j_2k_1}&...&\rho_{i_mj_mk_1,i_mj_mk_m}  \\
      \vdots & \vdots & \vdots & \vdots & \vdots&\vdots  \\
    \rho_{i_mj_mk_m,i_1j_1k_1}& ... & \rho_{i_mj_mk_m,i_1j_1k_m} & \rho_{i_mj_mk_m,i_1j_2k_1}&... &\rho_{i_mj_mk_m,i_mj_mk_m}  \\
  \end{pmatrix},\quad
\end{eqnarray}
\end{widetext}
where $i_1<...<i_m$, $j_1<...<j_m$ and $k_1<...<k_m$ with subindices
$i_1,...,i_m$ associated with the space $H_{A_1}$, $j_1,...,j_m$
with the space $H_{A_2}$ and $k_1,...,k_m$ with the space $H_{A_3}$.

\begin{theorem}\label{T1.3}
For any $N\otimes N \otimes N$ $(N\geq2)$ tripartite mixed quantum
state $\rho$, the concurrence $C(\rho)$ satisfies
\be\label{1.6}
C^2(\rho) \geq  c_{m\otimes m\otimes m}\sum_{P_{m\otimes
m\otimes m}} C^2(\rho_{m\otimes m\otimes m}),
\ee
where $m\geq2$, $c_{m\otimes m\otimes m}= \left[{N-1 \choose
m-1}\right]^{-1}\left[ {N-2 \choose m-2}\right]^{-2}$, and
$\sum_{P_{m\otimes m\otimes m}}$ stands for summing over all
possible $m\otimes m\otimes m$ mixed sub-states.
\end{theorem}

[Proof]. For any $N\otimes N \otimes N$ tripartite pure state $|\psi\rangle=\sum_{i=1}^{N}\sum_{j=1}^{N}\sum_{k=1}^{N}a_{ijk}\vert
ijk\rangle$, and any given term
\begin{eqnarray}\label{given term }
|a_{i_0j_0k_0}a_{p_0q_0m_0}-a_{i_0j_0m_0}a_{p_0q_0k_0}|^2,~~~k_0\neq m_0,
\end{eqnarray}
in Eq. (\ref{pure multi concurrence1}), if $i_0\neq p_0$ and $j_0\neq q_0$, then there are $\left[ {N-2 \choose m-2}\right]^{3}$ different $m\otimes m \otimes m$ sub-states $|\psi\rangle_{m\otimes m \otimes m}=B_1\otimes B_2\otimes B_3 |\psi\rangle$, with $B_1=|i_0\rangle\langle i_0|+|p_0\rangle\langle p_0|+\sum_{i=i_3}^{i_m} |i\rangle\langle i|$, $B_2=|j_0\rangle\langle j_0|+|q_0\rangle\langle q_0|+\sum_{j=j_3}^{j_m} |j\rangle\langle j|$, $B_3=|k_0\rangle\langle k_0|+|m_0\rangle\langle m_0|+\sum_{k=k_3}^{k_m} |k\rangle\langle k|$, where $\{|i\rangle\}_{i=i_3}^{i_m} \subseteq \{|i\rangle\}_{i=1}^N$, $\{|j\rangle\}_{j=j_3}^{j_m} \subseteq \{|j\rangle\}_{j=1}^N$, $\{|k\rangle\}_{k=k_3}^{k_m} \subseteq \{|k\rangle\}_{k=1}^N$, such that the term (\ref{given term }) appears in the concurrence of $|\psi\rangle_{m\otimes m \otimes m}=B_1\otimes B_2\otimes B_3 |\psi\rangle$. If $i_0= p_0$ and $j_0\neq q_0$, then there are $ {N-1 \choose
m-1}\left[ {N-2 \choose m-2}\right]^{2}$ different $m\otimes m \otimes m$ sub-states $|\psi\rangle_{m\otimes m \otimes m}=D_1\otimes B_2\otimes B_3 |\psi\rangle$, with $D_1=|i_0\rangle\langle i_0|+\sum_{i=i_2}^{i_m} |i\rangle\langle i|$, $\{|i\rangle\}_{i=i_2}^{i_m} \subseteq \{|i\rangle\}_{i=1}^N$, such that the term (\ref{given term }) appears in the concurrence of $|\psi\rangle_{m\otimes m \otimes m}=D_1\otimes B_2\otimes B_3 |\psi\rangle$. Otherwise, if $i_0\neq p_0$ and $j_0= q_0$, then there are $ {N-1 \choose
m-1}\left[ {N-2 \choose m-2}\right]^{2}$ different $m\otimes m \otimes m$ sub-states $|\psi\rangle_{m\otimes m \otimes m}=B_1\otimes D_2\otimes B_3\otimes |\psi\rangle$ with $D_2=|j_0\rangle\langle j_0|+\sum_{j=j_2}^{j_m} |j\rangle\langle j|$, $\{|j\rangle\}_{j=j_2}^{j_m} \subseteq \{|j\rangle\}_{j=1}^N$, such that the term (\ref{given term }) appears in the concurrence of $|\psi\rangle_{m\otimes m \otimes m}=B_1\otimes D_2\otimes B_3\otimes |\psi\rangle$. Since ${N-2 \choose m-2}\geq {N-1 \choose m-1}$ generally, the concurrences of
the pure state $\vert\psi\rangle$ and the sum of the concurrence of all the sub-states
$|\psi\rangle_{m\otimes m \otimes m}$ with respect to $\vert\psi\rangle$ have the following relation,
\begin{eqnarray}
C^2(\vert\psi\rangle)
\geq c_{m \otimes m\otimes m}\sum_{P_{m\otimes m\otimes m}}C^2(|\psi\rangle_{m\otimes m \otimes m}).
\end{eqnarray}

Therefore for mixed state $\rho=\sum p_i |\psi_i\rangle \langle \psi_i|$, we have
\begin{eqnarray*}
\ba{rcl}
&&C(\rho)\\&=&\displaystyle\min \sum_i p_i C(|\psi_i\rangle)\\[1mm]
&\geq& \displaystyle\sqrt{c_{m\otimes m\otimes m}}\min \sum_i p_i \left(\sum_{P_{m\otimes m\otimes m}}C^2(|\psi_i\rangle_{m\otimes m\otimes
m})\right)^{\frac{1}{2}}\\[1mm]
&\geq&\displaystyle\sqrt{c_{m\otimes m\otimes m}} \min \left[\sum_{P_{m\otimes m\otimes m}} \left(\sum_i p_i C(|\psi_i\rangle_{m\otimes m\otimes m})
\right)^2\right]^{\frac{1}{2}}\\[1mm]
&\geq&\displaystyle \sqrt{c_{m\otimes m\otimes m}}\left[\sum_{P_{m\otimes m\otimes m}} \left(\min \sum_i p_i C(|\psi_i\rangle_{m\otimes m\otimes m})
\right)^2\right]^{\frac{1}{2}}\\[1mm]
&=&\displaystyle \sqrt{c_{m\otimes m\otimes m}} \left[\sum_{P_{m\otimes m\otimes m}} C^2(\rho_{m\otimes m\otimes m})\right]^{\frac{1}{2}},
\ea
\end{eqnarray*}
where the relation $(\sum_j (\sum_i x_{ij})^2 )^{\frac{1}{2}} \leq
\sum_i (\sum_j x_{ij}^2)^{\frac{1}{2}}$ has been used in the second
inequality, the first three minimizations run over all possible pure
state decompositions of the mixed state $\rho$, while the last
minimization runs over all $m\otimes m \otimes m$ pure state decompositions
of $\rho_{m\otimes m \otimes m}=\sum_i p_i |\phi_i\rangle \langle
\phi_i|$ associated with $\rho$. \qed

(\ref{1.6}) gives a lower bound of $C(\rho)$. One can estimate $C(\rho)$ by calculating the concurrence of the
sub-states $\rho_{m\otimes m\otimes m}$, $2\leq m\leq N-1$. Different choices of $m$ may give rise to
different lower bounds. A convex combination of these lower bounds is still a lower bound. Hence generally we have
\begin{corollary}
For any $N\otimes N \otimes N$ tripartite mixed quantum state $\rho$, the concurrence $C(\rho)$ satisfies
\begin{eqnarray}
C^2(\rho)\geq  \sum_{m=2}^{N} p_m\,c_{m\otimes m\otimes m}\sum_{P_{m\otimes m\otimes m}}C^2(\rho_{m\otimes m\otimes m}),
\end{eqnarray}
where $0\leq p_m\leq1$ and $\sum_{m=2}^{N} p_m=1$.
\end{corollary}

The lower bound (\ref{1.6}) is in general not operationally computable, as we still have no analytical results
for concurrence of lower dimensional states. Nevertheless, we have already some analytical lower bounds for three-qubit mixed
quantum states \cite{Xiu-hong Gao.2006}. If we replace the computation of concurrence of lower dimensional sub-states
``$\rho_{m\otimes m\otimes m}$" by that of the lower bounds of these sub-states, (\ref{1.6}) gives an
operational lower bound based on known lower bounds. The lower bound obtained in this way should be the same or better than
the previous known lower bounds. Hence (\ref{1.6}) may be used to improve the existing lower bounds in this sense.
We first present an operational analytical lower bound for three-qubit mixed quantum states.

\begin{theorem}\label{th lowerbound from neg}
The concurrence $C(\rho)$ of three-qubit  mixed quantum state $\rho$ satisfies
\begin{eqnarray}\label{th three paritite nnn}
C^2(\rho)\geq\sum_{j=1}^3\max[(||\rho^{\mathcal{T}_j}||-1)^2,\, (||R_{j,\bar{j}}(\rho)||-1)^2],
\end{eqnarray}
where $\rho^{\mathcal{T}_{j}}$ stands for the partial transposition of $\rho$ with respect to
the $j$-th subsystem $A_j$, $R_{j,\bar{j}}(\rho)$ is the realignment of
$\rho$ with respect to the bipartite partition between $j$-th and the rest systems,
$||A||=Tr\sqrt{AA^\dagger}$ is the trace norm of a matrix.
\end{theorem}

[Proof]. For three-qubit state $|\psi\rangle$,  one has,
\begin{eqnarray*}
&&1-Tr \rho_{A_j}^2\\&&=\frac{1}{2}(\lVert(|\psi\rangle \langle \psi|)^{\mathcal{T}_{j}}\rVert-1)^2\\
&&=\frac{1}{2}(\lVert R_{j,\bar{j}}(|\psi\rangle \langle \psi|)\rVert-1)^2.
\end{eqnarray*}
According to (\ref{cpure}) we obtain
\begin{eqnarray*}
&&C^2(|\psi\rangle)\\&&=\sum_{j=1}^3(\lVert(|\psi\rangle \langle \psi|)^{\mathcal{T}_{j}}\rVert-1)^2\\
&&=\sum_{j=1}^3(\lVert R_{j,\bar{j}}(|\psi\rangle \langle \psi|)\rVert-1)^2.
\end{eqnarray*}

Assume that $\sum_ip_i|\psi_i\rangle \langle \psi_i|$ is the optimal
decomposition of $\rho$ achieving the infimum of $C(\rho)$. Then
\be \ba{rcl}
C^2(\rho)&=&\displaystyle\left(\sum_ip_iC(|\psi_i\rangle)\right)^2\\
&=& \left(\sum_ip_i\sqrt{\sum_j( \lVert(|\psi_i\rangle \langle \psi_i|)^{\mathcal{T}_{j}}\rVert-1)^2}\right)^2\\[1mm]
&\geq&\displaystyle\sum_{j=1}^3\left(\sum_i(p_i\lVert(|\psi_i\rangle \langle \psi_i|)^{\mathcal{T}_{j}}\rVert-1)\right)^2\\
&\geq&\sum_{j=1}^3\left(\lVert\rho^{\mathcal{T}_j}\rVert-1\right)^2.
\ea
\ee
Similarly one can prove that $C^2(\rho)\geq\sum_{j=1}^3(||R_{j,\bar{j}}(\rho)||-1)^2$.
\qed

To see the tightness of the inequality (\ref{th
three paritite nnn}), we consider the following example.

{\it Example 1.} Let us consider the D\"{u}r-Cirac-Trarrach state \cite{W.D},
\begin{equation}\label{1.8}
\begin{array}{rcl}
\rho_{DCT}=&&\sum_{\sigma=\pm}\lambda^{\sigma}_0\vert\psi^{\sigma}_0\rangle\langle\psi^{\sigma}_0\vert+\sum_{j=1}^3 \lambda_j (\vert\psi^{+}_j\rangle\langle\psi^{+}_j\vert\\[2mm]
&&+\vert\psi^{-}_j\rangle\langle\psi^{-}_j\vert),
\end{array}
\end{equation}
where $\vert\psi_0^{\pm}\rangle=\frac{1}{\sqrt{2}}(\vert
000\rangle\pm\vert 111\rangle)$, $$|\psi^\pm_j\rangle=\frac{1}{\sqrt{2}}(|j\rangle_{AB}|0\rangle_C\pm|(3-j)\rangle_{AB} |1\rangle_C),$$ $|j\rangle_{AB}=|j_1\rangle_A |j_2\rangle_B$ with $j=j_1j_2$ in binary notation. From our lower bound (\ref{th
three paritite nnn}), we obtain $C(\rho_{DCT})\geq\sqrt{\frac{4}{27}}\approx
0.385$ for $ \lambda^+_0=\frac{1}{6},\lambda^-_0=\frac{1}{2}$ and
$\lambda_{1}=\lambda_{2}=\lambda_{3}=\frac{1}{18}$.

Another lower bound of concurrence for three-qubit mixed quantum states had been given in
Ref. \cite{Xiu-hong Gao.2006}. Form Ref. \cite{Xiu-hong Gao.2006} the lower bound of concurrence for $\rho_{DCT}$ is
$C(\rho_{DCT})\geq {0.314}$, where the difference of a constant factor $\sqrt{2}$ in defining
the concurrence for pure states has been already taken into account.
Therefore the lower bound (\ref{th three paritite nnn}) is better than the
lower bound in Ref. \cite{Xiu-hong Gao.2006} in detecting entanglement of
the three-qubit mixed state $\rho_{DCT}$.

By using the analytical lower bounds (\ref{th three paritite nnn}) for three-qubit quantum states, from (\ref{1.6})
we have

\begin{corollary}
For any $N\otimes N \otimes N$ tripartite mixed quantum state $\rho$, the concurrence $C(\rho)$ satisfies
\begin{widetext}
\begin{eqnarray}\label{c2}
C^2(\rho)\geq  \frac{1}{N-1} \sum_{P_{2\otimes 2\otimes
2}}\max\left[\sum_{j=1}^3(||\rho_{2\otimes 2\otimes
2}^{\mathcal{T}_j}||-1)^2,\,\sum_{j=1}^3(||R_{j,\bar{j}}(\rho_{2\otimes 2\otimes
2})||-1)^2\right].
\end{eqnarray}
\end{widetext}
\end{corollary}

(\ref{1.6}) presents a lower bound of concurrence for $N\otimes N \otimes N$ tripartite mixed quantum states.
Generally it is not operational, while (\ref{th three paritite nnn}) gives an operational
lower bound of concurrence. The combination of these two results
gives rise to operational lower bounds for general $N\otimes N \otimes N$ states.

{\it Example 2.} We consider the $3\otimes 3\otimes 3$ state
$\rho=\frac{1-x}{27}I_{27}+x\vert\psi^+\rangle\langle\psi^+\vert$,
where $0\leq x\leq1$ represents the degree  of the depolarization,
$\vert\psi^+\rangle=\frac{1}{\sqrt{2}}(\vert000\rangle+\vert222\rangle).$
As $\rho^{\mathcal{T}_{\mathcal{Y}_i}}=(\rho^{\mathcal{T}_{\mathcal{Y}_i}})^{\dagger} $, the square root of
the eigenvalues of $\rho^{\mathcal{T}_{\mathcal{Y}_i}}(\rho^{T_{\mathcal{Y}_i}})^{\dagger}$ is the
absolute value of the eigenvalues of $\rho^{\mathcal{T}_{\mathcal{Y}_i}}.$ According to the
above Corollary, our result can detect the entanglement of $\rho$ when $\frac{2}{29}\leq x \leq 1$, see
Fig.1.

\begin{figure}[htpb]
\renewcommand{\captionlabeldelim}{.}
\renewcommand{\figurename}{Fig.}
\centering
\includegraphics[width=7cm]{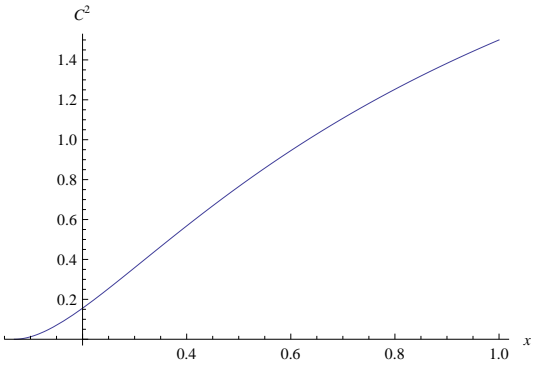}
\caption{{\small The lower bound concurrence of $\rho$ for $\frac{2}{29}\leq x\leq1$.}}
\label{2}
\end{figure}

Now we generalize our results to  arbitrary dimensional $n$-partite
systems. Let $H_{A_1} $, $H_{A_2}$, ..., $H_{A_n} $ be $N$
dimensional vector spaces respectively. A pure state
$\vert\psi\rangle\in H_{A_1}\otimes H_{A_2}\otimes \cdots\otimes
H_{A_n}$ has the form,
\begin{equation}
\vert\psi\rangle=\sum_{i_1=1}^{N}\sum_{i_2=1}^{N}...\sum_{i_n=1}^{N}a_{i_1i_2...i_n}\vert i_1i_2...i_n\rangle,
\end{equation}
where $a_{i_1i_2...i_n}\in \Cb$,
$\sum_{i_1i_2...i_n}|a_{i_1i_2...i_n}|^2=1, \{\vert
i_1i_2...i_n\rangle\}$ is the  basis of $H_{A_1}\otimes
H_{A_2}\otimes...\otimes H_{A_n}$. The concurrence of $\vert\psi\rangle$ has the form \cite{anote},
\be
\label{pure multi concurrence}
C(\vert\psi\rangle)=\sqrt{\sum_p\sum_{\alpha,\acute{\alpha},\beta,\acute{\beta}}
|a_{\alpha\beta}a_{\acute{\alpha}\acute{\beta}}-a_{\alpha\acute{\beta}}a_{\acute{\alpha}\beta}|},
\ee
where $\sum_p$ stands for the summation over all possible
combinations of the indices of $\alpha, \beta$. $\alpha$ (or
$\acute{\alpha}$) and $\beta$ (or $\acute{\beta}$) span the whole
space of a given subindex of $a$. The concurrence is extended to
mixed state $\rho$ by the convex roof $C(\rho)=\min \sum_i p_i
C(|\psi_i\rangle)$ for all possible ensemble realizations $\rho=\sum
p_i |\psi_i\rangle \langle \psi_i|$.

For a given $N\otimes N \otimes \cdots
\otimes N$ pure state, we define its ``$m\otimes m\otimes \cdots \otimes m$"
pure states
\begin{eqnarray*}
|\psi\rangle_{m\otimes m\otimes \cdots \otimes m}
&&=\sum_{ {i_1=j_1}}^{j_{m}}\sum_{ {i_2=k_1}}^{k_{m}}\cdots\sum_{ {i_n=l_1}}^{l_{m}}
a_{{i_1} {i_2}\cdots  {i_n}} | {i_1} {i_2}\cdots {i_n}\rangle\\
&&=B_1\otimes B_2
\otimes \cdots \otimes B_n |\psi\rangle,
\end{eqnarray*}
where $B_1=\sum_{ {i_1=j_1}}^{j_{m}}|{i_1}\rangle\langle  {i_1}|$,
$B_2=\sum_{ {i_2=k_1}}^{k_{m}}|{i_2}\rangle\langle  {i_2}|$, ...,
$B_n=\sum_{ {i_n=l_1}}^{l_{m}} |{i_n}\rangle\langle  {i_n}|$, $\{j_1,\cdots,j_m\}\subseteq \{1,\cdots, N\}$, $\{k_1,\cdots,k_m\}\subseteq \{1,\cdots, N\}$, $\cdots$, and $\{l_1,\cdots,l_m\}\subseteq \{1,\cdots, N\}$,.
For a mixed state $\rho$, correspondingly we define
its ``$m\otimes m\otimes \cdots \otimes m$" sub-states
$$
\rho_{m\otimes m\otimes \cdots \otimes m}=B_1\otimes B_2
\otimes \cdots \otimes B_n\, \rho\, B_1^\dagger\otimes B_2^\dagger
\otimes \cdots \otimes B_n^\dagger.
$$
The concurrence
of $\rho_{m\otimes m\otimes \cdots \otimes m}$ is defined by
$C(\rho_{m\otimes m\otimes \cdots \otimes m})\equiv\min \sum_i p_i\,
C(|\phi_i\rangle$,
minimized over all possible $m\otimes m\otimes \cdots \otimes m$ pure
state decompositions of $\rho_{m\otimes m\otimes \cdots \otimes m}
=\sum_i p_i \,|\phi_i\rangle \langle \phi_i|$, with $\sum_i
p_i=tr(\rho_{m\otimes m\otimes \cdots \otimes m})$.
Similar to the tripartite case, we can prove the following theorem:

\begin{theorem}\label{th general multi lower bound}
For any $n$-partite $N$ dimensional mixed state
$\rho\in H_{A_1}\otimes H_{A_2}\otimes \cdots\otimes H_{A_n}$,
\begin{equation}
C^2(\rho)\geq c_{m\otimes m\otimes \cdots \otimes m}
\sum_{P_{m\otimes m\otimes \cdots \otimes m}}
C^2(\rho_{m\otimes m\otimes \cdots \otimes m}),
\end{equation}
where $c_{m\otimes m\otimes \cdots \otimes m}$ is a fixed number
depending on $m$,
$\sum_{P_{m\otimes m\otimes \cdots \otimes m}}$ stands for summing over
all possible $m\otimes m\otimes \cdots \otimes m$ mixed states.
\end{theorem}

\section{Conclusions}

In summary, we have proposed a method in constructing hierarchy lower
bounds of concurrence for tripartite mixed states, in terms of the
concurrences of all the lower dimensional mixed sub-states.
The lower bounds may be used to improve all the existing lower
bounds of concurrence. The approach can be readily generalized to arbitrary
dimensional multipartite systems.

\noindent{\bf Acknowledgments}\, 
This work is supported by PHR201007107.

\end{document}